\documentclass[aps,showpacs]{revtex4}
\usepackage{amssymb,amsmath}
\usepackage{graphicx}
\usepackage{dcolumn}
\newcommand{\be}{\begin{equation}}
\newcommand{\ee}{\end{equation}}
\newcommand{\bea}{\begin{eqnarray}}
\newcommand{\eea}{\end{eqnarray}}
\newcommand{\ds}{\displaystyle}

\newcommand{\vk}{{\mathbf{k}}}
\newcommand{\vp}{{\mathbf{p}}}
\newcommand{\p}{\partial}
\newcommand{\bc}{\begin{center}}
\newcommand{\ec}{\end{center}}
\newcommand{\nn}{\nonumber}
\newcommand{\lb}{\label}
\newcommand{\re}[1]{(\ref{#1})}

\def\trans{\mbox{\tiny$\bot$}} 
\def\longi{\mbox{\tiny$\|$}}   
\begin{document}

\title{
Vacuum creation of quarks at the time scale of QGP thermalization
and strangeness enhancement in heavy-ion collisions
   }
\author{A.V.~Prozorkevich}
\affiliation{Physics Department, Saratov State University,
RU-41026 Saratov, Russia
\vspace*{1ex}}
\author{S.A.~Smolyansky}
\affiliation{Physics Department, Saratov State University,
RU-41026 Saratov, Russia
\vspace*{1ex}}
\author{V.V.~Skokov}
\affiliation{Bogoliubov Laboratory of Theoretical Physics, Joint 
Institute for Nuclear Researches, RU-141980 Dubna, Russia
\vspace*{1ex}}
\author{E.E.~Zabrodin}
\affiliation{Centre of Mathematics for Applications, University of 
Oslo, N-0316 Oslo, Norway
\vspace*{1ex}}
\affiliation{Institute for Nuclear Physics, Moscow
State University, RU-119899 Moscow, Russia
\vspace*{1ex}}

\date{\today}

\begin{abstract}
The vacuum parton creation in quickly varying external fields is
studied at the time scale of order 1\,fm/$c$ typical for the
quark-gluon plasma formation and thermalization. To describe the
pre-equilibrium evolution of the system the transport kinetic equation
is employed. It is shown that the dynamics of production process at
times comparable with particle inverse masses can deviate considerably
from that based on classical Schwinger-like estimates for homogeneous
and constant fields. One of the effects caused by non-stationary
chromoelectric fields is the enhancement of the yield of $s\bar{s}$
quark pairs. Dependence of this effect on the shape and duration of
the field pulse is studied together with the influence of string 
fusion and reduction of quark masses.
\end{abstract}

\maketitle

\section{Introduction}
\lb{intro}

Multiple particle production in relativistic heavy-ion collisions is
not fully understood yet. One of the most popular microscopic
approaches to this phenomenon is formulated within the chromoelectric
flux tube (or rather string) model \cite{CNN79}. The flux tubes are
assumed to be initially stretched between the constituents (quarks
and diquarks) of the nucleons of colliding nuclei. As the constituents
are flying away, the energy of a color tube increases making it
unstable against the production of $q\bar{q}$ or $d\bar{d}$ pairs
from the vacuum. New hadrons, created in the course of the color tube
fragmentation, can also interact within hot and dense nuclear matter,
and the in-medium cascade develops.

The formation and subsequent break-up of color tubes/strings is a
common feature adapted by all microscopic string models (see, e.g.,
\cite{lund,Kaid99,qgsm,venus,urqmd} and references therein), which are
intended to describe dynamics of heavy-ion collisions at relativistic
energies. For a uniform chromoelectric field $E$ the probability to
create a pair of quarks with mass $m$, effective charge $e$, and
transverse momentum $p_t$ per unit time and per unit volume reads
\cite{CNN79}
\be \ds
P(p_t) d^2p_t = - \frac{|eE|}{4\pi^3} \ln \left\{ 1 -
\exp\left[ -\frac{ \pi (m^2 + p_t^2)}{|eE|} \right] \right\}d^2p_t.
\lb{eq1}
\ee
The integrated probability is given by series expansion
\be \ds
P_m = \frac{(eE)^2}{4\pi^3} \sum \limits_{n=1}^\infty \frac{1}{n^2}
\exp \left( - \frac{\pi m^2 n}{|eE|} \right),
\lb{eq2}
\ee
which reproduces the classical Schwinger result \cite{schw} derived in
spinor quantum electrodynamics (QED) for $e^+e^-$ production rate in
the constant electric field. Usually, only the leading term in
Eq.~(\ref{eq2}) is taken into account. According to this formula, the
ratio of production rates of strange to nonstrange quark pairs,
widely known as strangeness suppression factor $\gamma_s$, is
\be \ds
\gamma_s = \frac{P(s\bar s)}{P(q\bar q)} =
\exp \left[ - \frac{\pi (m_s^2 -m_q^2)}{\kappa} \right],
\lb{eq3}
\ee
where $\kappa = |eE|$ is the so-called string tension. It appears
that microscopic models underestimate the yield of strangeness in
ultrarelativistic heavy-ion collisions \cite{low_str}. This issue is
very important, because the abundant yield of strange particles
was predicted \cite{raf91} as one of the signals of quark-gluon
plasma (QGP) creation. According to Eq.~(\ref{eq3}), the strangeness
production can be enhanced either by taking
into account string-string interaction, which leads to fusion of
strings and formation of color ropes with larger effective string
tension \cite{bnk84,sor92,am93,soff99}, or by dropping the quark
masses \cite{blei00}, e.g., due to chiral symmetry restoration.
Also, the effects related to the finiteness of the strings can
modify the production rates \cite{WW88,fbs}.

In the present Letter we explore another possibility: The
real fields emerging in heavy-ion collisions act at the time interval
comparable with the Compton scale. Since the dynamics of particle
creation in {\it time-dependent\/} homogeneous fields differs from
that of stationary fields \cite{grib}, it is essential to properly
modify the system description at early stage of nuclear collisions.
The kinetic equation (KE) is a convenient tool to study the
nonequilibrium evolution processes. The source term describing
the vacuum pair creation process can be incorporated into the KE
either in a phenomenological manner \cite{gatoff,asakawa} on the
basis of Schwinger-like formula, or derived in a more sophisticated
manner from the microscopic equations of motion \cite{smol}.
Although it is believed that the phenomenological source term
correctly reproduces qualitative features of the quantum mean field
theory \cite{kluger98}, such approximation has yet to be verified
for, e.g., time dependent fields or multicomponent systems. Compared
to the phenomenological treatment, the approach within the framework
of a transport equation \cite{smol} contains several new dynamical
aspects, such as longitudinal momentum dependence of the
distribution functions and non-Markovian character of the time
evolution. It takes into account effects of the field switching and
statistics, as well.
Therefore, the abundances of newly produced particles
may considerably deviate from the values obtained for the constant and
infinite field. The appearance of the non-Schwinger regime in a
creation of electron-positron pairs in periodic laser field has been
discussed in \cite{gan,pop,rob}. Another noteworthy feature of the
modified source term is the suppression of zero momentum bosons
\cite{smol1}. This circumstance causes the ``fermion dominance''
effect at the short time scales \cite{skok1} and can lead to the
abundant production of heavy fermions at expense of light bosons.

In case of the QGP creation the characteristic time of the field
variation is estimated to be of order of few fm/$c$ \cite{heinz} and,
therefore, the assumption of the space-time unvarying field is too
crude. Hence, it is necessary to elaborate on the dynamics of
particle vacuum production at short time scales, which are compatible
with the particle inverse masses. Particularly, the dependence of the
production rates on the shape and duration of the field pulse should
be investigated.

The paper is organized as follows. Kinetic equation with the
source term, describing the vacuum production of fermions, is derived
in the collisionless limit in Sec.~\ref{sec2}. Section~\ref{sec3}
presents study of the influence of the field pulse characteristics on
the time-dependent vacuum production rates for the energies of RHIC
and LHC. Comparison with the production rates, obtained for the
infinite and constant chromoelectric field by the classical
Schwinger-like formula, is also performed. Finally, conclusions are
drawn in Sec.~\ref{sec4}.

\section{Kinetic equation }
\lb{sec2}

As was mentioned above, the kinetic equation is a standard tool to
describe the non-equilibrium evolution of a many body system. The
general form of the KE for the distribution function $f(\vp,t)$ in
a strong spatially homogeneous time-dependent field is
\cite{gatoff}
\be \ds \lb{ke}
 \frac{\p f(\vp,t)}{\p t} + e\mathbf{E}(t) \frac{\p f(\vp,t)}{\p
\vp}= S(\vp,t)+ C(\vp,t),
\ee
where $S(\vp,t)$ and $C(\vp,t)$ are the source term and the collision
integral, respectively. The source term describes the vacuum
production of particle-antiparticle pairs in the external field, while
the collision term governs their rescattering dynamics, which drives
the system toward thermal equilibrium. For the sake of clarity, it is
relevant to consider the collisionless approximation, where the
collision term in the r.h.s. of Eq.~(\ref{ke}) is omitted.

The kinetic description of quarks in framework of the Wigner function
formalism \cite{oh,qcd} leads to a very complicated system of partial
differential equations hard to solve both numerically and analytically.
Therefore, following the approach of \cite{grib,smol}, we employ the
canonical Bogoliubov transformation method within the QED approximation.
To derive the source term for the system of fermions in an external
electric field, we start from the Dirac equation
\be \ds \lb{eq5}
(i\gamma^\mu\partial_\mu-e\gamma^\mu A_\mu-m)\psi(x)=0\,.
\ee
Using the simple field configuration with vector potential in the
Hamilton gauge $A^\mu=(0,0,0,A(t))$ and homogeneous electric field
$\mathbf{E}(t)=(0,0,E(t))$, $E(t)=-\dot A(t)$, one looks for the
solutions of the Eq.~(\ref{eq5}) in the form
\be \ds \lb{eq6}
\psi^{(\pm)}_{\vk r}(x) = \left[i\gamma^0\partial_0-\mathbf{\gamma k}
+e\gamma^3A(t)+m\right] \ \chi^{(\pm)}(\vk,t) \ R_r \
{\rm e}^{i\vk\bar x},
\ee
where the superscript $(\pm)$ denotes eigenstates with the positive
and negative frequencies.  Herein the spinors $R_r$ $(r = 1,2)$ are
eigenvectors of the matrix $\gamma^0\gamma^3$ satisfying the
condition  $R^+_r R_s = 2\delta _{rs} \,.$ The functions
$\chi^{(\pm)}(\vk,t)$ obey the oscillator-type equation
\be \ds \lb{eq7}
{\ddot \chi}^{(\pm)}(\vk,t)+\big[\omega^2(\vk,t)-ie{\dot A}(t)\big] \
\chi^{(\pm)}(\vk,t)=0\,\,,
\ee
where we define  the total energy
$\omega^2(\vk,t)=\varepsilon_{\trans}^2+p_{\longi}^2(t)$,
the transverse energy $\varepsilon_{\trans}^2=m^2+\vk^2_{\trans}$,
and the longitudinal momentum $p_{\longi}(t)=k_{\longi}-eA(t)$. The
solutions $\chi^{(\pm)}(\vk,t)$ of Eq.~(\ref{eq7}) for positive and
negative frequencies are fixed by their asymptotic behavior at
$t_0=t\rightarrow -\infty$, where $\dot A(t_0) = 0$. In this limit, the field
operators $\psi(x)$ and ${\bar \psi}(x)$ can be decomposed
by the complete and orthonormalized set of spinor functions (\ref{eq6})
as follows:
\be \ds \lb{eq8}
\psi(x)=\sum\limits_{r,\vk} \left[\psi^{(-)}_{\vk r}(x)\ b_{\vk r}(t_0) +
\psi^{(+)}_{\vk r}(x) \ d^+_{-\vk r}(t_0) \right] \,\,.
\ee
The operators $b_{\vk r}(t_0),b^+_{\vk r}(t_0)$ and
$d_{\vk r}(t_0),d^+_{\vk r}(t_0)$
describe the annihilation and creation of particles and antiparticles
in the in-state $|0_{in}\rangle$
and obey the standard anticommutation rules. The time evolution
leads to the mixing of states with positive and negative energies
and, therefore, non-diagonal terms in the Hamiltonian corresponding
to Eq.~(\ref{eq5}) emerge. The diagonalization of the Hamiltonian,
which is equivalent to the transition to quasiparticle representation,
is performed by the time-dependent Bogoliubov transformation
\bea \ds \lb{eq9}
\begin{array}{lcccl} b_{\vk r}(t) &=& \alpha_\vk (t)\ b_{\vk r}(t_0)
&+& \beta_\vk (t)\ d^+_{-\vk r}(t_0)\ , \\
d_{\vk r}(t) &=& \alpha_{-\vk} (t) \
d_{\vk r}(t_0) &-& \beta_{-\vk}(t) \ b^+_{-\vk r}(t_0)\ , \end{array}
\eea
with the imposed condition $|\alpha_\vk(t)|^2+|\beta_\vk(t)|^2=1\,.$
The new operators $b_{\vk r}(t)$ and $d_{\vk r}(t)$ describe the
processes of quasiparticle creation and annihilation. By virtue of
the Lagrange multipliers, one can find from the equations of motion
\re{eq7} that the coefficients in the Bogoliubov transformation
\re{eq9} are connected via the relations \cite{grib}
\bea \ds \lb{eq10}
\begin{array}{lcl}  {\dot \alpha}_\vk(t) &=& \quad
{\ds\frac{eE(t)\varepsilon_{\trans}}{2\omega^2(\vk,t)}}\
\beta^*_\vk(t) \ {\rm e}^{2i\theta(\vk,t_0,t)}\,\,, \\ {\dot
\beta}^*_\vk(t)&=&-{\ds\frac{eE(t)\varepsilon_{\trans}}
{2\omega^2(\vk,t)}}\
\alpha_\vk(t) \ {\rm e}^{-2i\theta(\vk,t_0,t)}\,\,, \end{array}
\eea
where the dynamical phase is defined as
\be \ds \lb{eq11}
\theta(\vk,t_0,t) = \int^t_{t_0}dt'\omega(\vk,t')\,\,.
\ee
To absorb the dynamical phase it is convenient to introduce new
operators
\be \ds \lb{eq12}
\begin{array}{lcl}
B_{\vk r}(t) &=& b_{\vk r}(t) \  e^{-i\theta(\vk,t_0,t)}\,,\qquad \\
D_{\vk r}(t) &=& d_{\vk r}(t)\  e^ {-i\theta(\vk,t_0,t)},
\end{array}
\ee
which obey the anti-commutation relations:
\be \ds \lb{eq13}
\{B_{\vk r}(t),B^+_{\vk ' r'}(t)\}=\{D_{\vk r}(t),D^+_{\vk ' r'}(t)\}=
\delta_{rr'} \ \delta_{\vk \vk '} \, .
\ee
These operators satisfy the Heisenberg-type equations of motion
\bea \ds \lb{eq14}
\begin{array}{lcl} {\ds\frac{dB_{\vk r}(t)}{dt}=-\frac{e
E(t)\varepsilon_{\trans}}{2\omega^2(\vk,t)}} \ D^+_{-\vk r}(t) +
i \ [H(t), \ B_{\vk r}(t)]\,\,,
 \\ {\ds\frac{dD_{-\vk r}(t)}{dt}=\phantom{-}
\frac{e E(t)\varepsilon_{\trans}}
{2\omega^2(\vk,t)}}\ B^+_{\vk r}(t)+i \ [H(t),\ D_{-\vk r}(t)] \ ,
\end{array}
\eea
where $H(t)$ is the Hamiltonian of the system of quasiparticles
\be \ds \lb{eq15}
H(t)=\sum_{r,\vk} \omega(\vk,t)\left[B^+_{\vk r}(t) \ B_{\vk r}(t)-
D_{-\vk r}(t) \ D^+_{-\vk r}(t)\right]\,\, .
\ee
The first term in the r.h.s. of Eqs.~(\ref{eq14}) arises because of
the unitary non-equivalence of the transition from the
representation \re{eq8} to the quasiparticle one.

Next consider the evolution of the distribution function of
quasiparticles with the momentum $\vk$ and spin $r$ defined as
\be \ds \lb{eq16}
f_r(\vk,t) =\, \langle 0_{in}|b^+_{\vk r}(t) \ b_{\vk r}(t)|0_{in}
\rangle\, =\,
\langle 0_{in}|B^+_{\vk r}(t) \ B_{\vk r}(t)|0_{in} \rangle \ .
\ee
According to the charge conservation the distribution functions for
particles and anti-particles are related as $f_r(\vk,t) = {\bar
f}_r(-\vk,t)$. Taking derivative in Eq.~(\ref{eq16}) with respect
to time $t$ we have
\be \ds \lb{eq17}
\frac{d f_r(\vk,t)}{dt}=
-\frac{eE(t)\,\varepsilon_{\trans}}{\omega^2(\vk,t)} \ {\rm
Re}\{\Phi_r(\vk,t)\} \ .
\ee
Here the function $\Phi_r(\vk,t)=\langle 0_{in}|D_{-\vk r}(t) \
B_{\vk r}(t)|0_{in} \rangle$ describes the vacuum production of pairs
in external electric field $E(t)$. Applying the equations of motion
(\ref{eq14}), one finds
\be\ds\lb{eq18}
\frac{d\Phi_r(\vk,t)}{dt} = \frac{e E(t)\,\varepsilon_{\trans}}
{2\omega^2(\vk,t)}\left[2f_r(\vk,t)-1\right]-
2i\omega(\vk,t) \ \Phi_r (\vk,t)\ .
\ee
The solution of Eq.~(\ref{eq18}) with the initial condition
$\Phi_r(\vk,t_0)=0 $ may be written in the following integral form
\be\ds\lb{eq19}
\Phi_r(\vk,t) = \frac{\varepsilon_{\trans}}{2}\int\limits_{t_0}^t dt'
\frac{e E(t')}{\omega^2(\vk,t')}\left[2f_r(\vk,t')-1\right]
\exp{[ 2i\theta(\vk,t',t)]}\,\,.
\ee
Inserting this result into the r.h.s. of Eq.~(\ref{eq17}) we obtain
the anticipated kinetic equation
\be\ds\lb{eq20}
\frac{df_r(\vk,t)}{dt} = \frac{e E(t)\varepsilon_{\trans}}
{2\omega^2(\vk,t)}\int\limits_{t_0}^t \! dt' \,
\frac{e E(t')\varepsilon_{\trans}}{\omega^2(\vk,t')}\left[
1-2f_r(\vk,t')\right]\cos{ [2\,\theta(\vk,t',t)]}\ .
\ee
Since the distribution function does not depend on spin, the
subscript $r$ can be dropped: $f_r \equiv f$. Substitution $\vp =
\vk - e \mathbf{A}(t)$, where the 3-momentum is decomposed onto the
transverse and longitudinal components $\vp =
\vp (p_{\trans}, p_{\longi}(t))$, yields to the reduction of the KE
(\ref{eq20}) to Eq.~\re{ke} with the source term
\be \ds \lb{source}
S(\vp,t) = \frac{e^2}{2}E(t)\mathrm{w}(\vp)
\!\int\limits_{t_0}^t \!dt_1 \ E(t_1)\mathrm{w}(\vp(t,t_1))
\left[1-2f(\vp(t,t_1),t_1) \right]\cos\left(2\!\int\limits^t_{t_1}dt_2 \
\omega(t,t_2) \right),
\ee
where $\mathrm{w}(\vp) = \varepsilon_{{\trans}}/\omega(\vp)$ and
\bea \lb{eq22}
\omega(t,t_1) &=&
\sqrt{\varepsilon_{{\trans}}^2+p_{\longi}^2(t,t_1)}, \nn \\
\vp(t,t_1) &=& \vp-e\int_{t_1}^t \!\mathbf{E}(t_2) dt_2 \ .
\eea

The source term \re{source} demonstrates several interesting
features, such as the dependence on particle longitudinal and
transverse momentum, the account for spin and statistics, and the
non-Markovian character of the time evolution. The memory effects are
caused by the time integration over the statistical factor $(1-2f)$
and the non-local cosine function \cite{smol1}, while the structure of
the coefficient $\mathrm{w}(\vp)$ defines the shape of the momentum
distribution of created particles.

In the collisionless approximation the kinetic equation \re{ke} with
the source term given by Eq.~\re{source} can be transformed into the
system of three ordinary differential equations \cite{mam}
\bea \ds \lb{eq23}
2\dot f&=& eE\mathrm{w}v_1\ , \nn \\
\dot v_1&=& eE\mathrm{w} (1-2f)-2\omega v_2\ ,\\
\dot v_2&=& 2\omega v_1\ , \nn
\eea
where the dot denotes the full time derivative, and the auxiliary
functions $v_1, v_2$ defined as
\be \ds \lb{eq24}
v_1=-2\,{Re}\Phi_r(\vk,t), \qquad
v_2=-2\,{Im}\Phi_r(\vk,t).
\ee

If the field strength is of order of the critical value, then it is
necessary to take into account the back reaction of produced particles
on primary field \cite{gatoff,KES}. The newly created particles
polarize the vacuum and are accelerated by the external field. Their
motion generates an internal field that in its turn modifies the
initial background field. For the description of this phenomenon, the
background field $E(t)$ in \re{source} should be replaced by the sum
$E(t)=E_{ex}(t)+E_{in}(t)$, where the generated internal field
$E_{in}$ can be found from the Maxwell equation \cite{smol}
\be \ds \label{eq25}
\dot E_{in}=- \frac{e g_f}{(2\pi)^3}\int \frac{d^3p}{\omega}\left[
2\,p_{\longi}\, f + p_{{\trans}}v_1\right]\ ,
\ee
with $g_f=2N_c$ being the degeneracy factor and $N_c$ the number of
color degrees of freedom ($N_c=3$).
The total current density in the r.h.s. of
Eq.~\re{eq25} is the sum of conductivity and vacuum polarization
currents, respectively. The integrand in \re{eq25} contains the
logarithmic divergence which should be removed somehow by means of a
regularization procedure \cite{mam}. We use here the simple
ultraviolet cut-off of momentum integration on the border of a grid.
The Eqs.~\re{eq23} and \re{eq25} represent the closed system of
equations for the numerical analysis of the back reaction problem.

\section{Vacuum creation of quarks with different masses}
\lb{sec3}

The derived formalism can be applied to study the vacuum creation of
quark-antiquark pairs in heavy ion collisions. Of special interest is
the analysis of effects caused by the fast change of
(chromo)electrical field at the time scale compatible with the
inverse quark masses. To investigate a role of the field switching
on/off effects, we approximate the time dependence of the flux-tube
field by a short pulse oriented along the collision axis of primordial
nuclei
\be \ds \lb{gauss}
E(t)=E_m\exp{[-(2t/\tau)^n]}\ ,
\ee
where $E_m$ is the field magnitude and $\tau$ is the effective pulse
width. The integer exponent $n$ governs the steepness of the pulse:
for $n\gg 1$ (we restrict ourselves to $n_{max}=6$), the pulse
\re{gauss} becomes close to the rectangular one. The case with $n=2$
reproduces well the soliton-like pulse
\be \ds \lb{eq27}
\begin{array}{lcl}
E(t) &=&  E_m\cosh^{-2}(2t/\tau)\ ,\\
A(t) &=& -E_m(\tau/2)\tanh(2t/\tau)\ ,
\end{array}
\ee
for which the Dirac equation allows exact analytical solution
\cite{grib,nar}.

It is convenient to define the time dependent strangeness suppression
factor $f_s$ as the ratio of densities of strange $(s)$ to nonstrange
$(q)$ quarks
\be \ds \lb{gammat}
f_s(t) = \frac{n(m_s,t)}{n(m_q,t)}\ ,
\ee
where in the case of axial symmetry
\be \ds \lb{dens}
n(m_f,t)=g_f \int\frac{d^3p}{(2\pi)^3} f(m_f,p_{\trans},p_{\longi})\ .
\ee
The system \re{eq23} is integrated by the
Runge-Kutta method with the zero initial conditions
$f(\vp,t_0) = v_1(\vp,t_0) = v_2(\vp,t_0) = 0$.
The momentum dependence of the distribution function
$f(p_{\longi}, p_{\trans}, t)$ is determined by coarse-graining of the
momentum space to a 2-dimensional grid; in each of its node the system
of equations \re{eq23} is solved. The parameters of the grid depend on
the field strength, the typical values are $\Delta p\approx 0.1m$
(step of the grid) and $p_{max}\approx 15-20m$ (boundary of the grid),
so the total number of the solved equations is about $10^6$.

{\bf Effect of the field pulse duration}.
To study the influence of the field pulse duration on the ratio of
strange and nonstrange quark pairs we compare results, obtained for
the time-dependent ratio \re{gammat} with the soliton-like pulse
\re{eq27}, with those yielded for the constant chromoelectric field
by Eq.~\re{eq3}. Input parameters, such as quark masses and string
tension, are chosen according to \cite{amel}, namely
\bea\lb{set1}
m_s &=& 350\, {\rm MeV}\ ,\nn\\
m_{u,d} &=& 230\, {\rm MeV}\ ,\nn\\
\kappa &=& |eE|=0.9\, {\rm GeV/fm}\ .
\eea
Since the critical field is defined as $eE_{cr}=m^2$, the chosen
value of the field is under-critical one for heavy strange quarks and
over-critical one for the light nonstrange quarks. To solve the
Maxwell equation \re{eq25} the value of the effective charge $e$
should be determined. This can be done on basis of the estimates for
the initial energy density $\varepsilon$, which varies from
50 GeV/fm$^3$ at RHIC to 520 GeV/fm$^3$ at LHC \cite{cooper}. Assuming
for simplicity that all initial energy is deposited in the field
sector, one obtains $E_{ex}\approx 3.16$ GeV/fm for RHIC and
$E_{ex}\approx 10$ GeV/fm for LHC. The corresponding values of the
effective charge are $e \approx 0.22$ and $e \approx 0.07$ for RHIC
and LHC, respectively.

The results of calculations of the asymptotic value $f_s(t\gg\tau)$
of the suppression factor $f_s$
are depicted in Fig.~\ref{f1} for the field pulse \re{eq27}.
Trivially, the Schwinger-like estimate for the constant and
infinite field \re{eq3} gives the constant value $f_s = 0.3$.
The time-dependent case is more complex: for pulses shorter than the
particle inverse masses the creation of $s\bar{s}$ quark pairs is
significantly enhanced. This is a direct consequence of the
uncertainty relation for the energy and time.
It is worth mentioning that already at $\tau \geq 3$ fm/$c$ the exact
result is very close to that given by the Schwinger formula \re{eq3}.
However, for the characteristic time of QGP formation $\tau_0 \sim 1$
fm/$c$ \cite{Bj83}, the production probability of strange quarks in
the soliton-like field yield \re{eq27} is at least 1.5 times larger
than that in the stationary case.

{\bf Influence of the field pulse shape.} Figure~\ref{f2} shows the
time evolution of the factor $f_s(t)$ during action of field pulses
\re{gauss} with $n=6$ and \re{eq27} with the same width $\tau=1$
fm/$c$. We see, that production of strange quarks increases with rise
of the power $n$ in exponent in Eq.\re{gauss}.
Note also, that the intermediate effective value of the time dependent
ratio $f_s$ is considerably larger than its final one: it varies from
about 0.6 to 0.39 for the soliton-like pulse and from 0.75 to 0.52 for
the rectangular one.
The evolution of densities of the created particles is presented in
Fig.~\ref{f3}. The density of both light and heavy quark pairs in the
pulsing field initially increases. As the field saturates and
starts to decrease, the process of particle absorption by the field
dominates the particle production one. The longer the field pulses,
the stronger absorption. Therefore, the way of the field oscillation
can significantly change the value of the suppression factor. The
similar result concerning the role of pulse shape of periodic laser
field on the electron-positron vacuum production rate was obtained in
\cite{pop} within the framework of the approximate imaginary time
method.

In Ref. \cite{soff99} two ways of increasing the strangeness
production within the framework of the Schwinger mechanism have been
discussed, namely (i) either the field strength (string tension) is
increasing, or, equivalently, (ii) quark masses are dropping due to
chiral symmetry restoration. Assuming a significant reduction of quark
masses from \re{set1} to the current quark masses $m_s=0.23$~GeV,
$m_q=0.01$~GeV \cite{soff99}, we get from Eq.~\re{eq3} the enhancement
of strangeness production in a constant infinite field from
$\gamma_s \approx 0.3$ to $\gamma_s
\approx 0.4$, which is equivalent to increase of the string tension
$\kappa$ from 0.9 GeV/fm to 1.22 GeV/fm.

Let us perform similar fitting procedure for the rectangular pulse
given by Eq.~\re{gauss}. The corresponding dynamical picture is
displayed in Fig.~\ref{f4}. It appears that the scenarios considered
above are not fully equivalent in dynamical sense, because the
production of light particles at very short times is apparently
suppressed.
Therefore, the reduction of quark masses due to the chiral symmetry
restoration is even more effective for the enhancement of strangeness
production than the mechanism of string overlap, which leads to
the formation of color rope and increase of the string tension.

The solution of Eqs.\re{eq23} and \re{eq25} shows that the back
reactions play minor role in the pair production processes with the
values of parameters given by the set \re{set1}, because the current
of secondaries is quite weak. The increase of the string tension
$\kappa$ from 0.9 GeV/fm to 1.22 GeV/fm or the reduction of quark
masses to $m_s=0.23$ GeV and $m_q=0.01$ GeV does not significantly
modify this scenario.

\section{Conclusions}
\lb{sec4}

Kinetic equation with the source term, describing the vacuum
production of fermions in a strong, time-dependent field, is used
for the two-component system investigation
in the collisionless approximation. It is shown that dynamics of
vacuum particle creation at time scale compatible with the inverse
mass of particle depends essentially on the specific characteristics
of field configuration (a form and a duration of the field pulse).
In particular, the Schwinger-like regime of particle creation might
not be realized at the typical time scale of QGP formation,
$\tau_0 \sim 1$ fm/$c$. As a consequence, production of heavy
strange quarks becomes more abundant. The role of string fusion
and reduction of quark masses due to the chiral symmetry restoration
which can alter the strangeness production is studied. Within the
proposed dynamical scenario the mass reduction mechanism appears to
be more effective for the enhancement of strangeness yield than the
formation of color rope. Since the current of the produced secondaries
is weak, the contribution of the back reactions to the production
process is small. Our study shows that the time evolution picture of
the gluon field should be incorporated consistently into the color
flux tube model for quantitative description of parton production in
ultrarelativistic heavy ion collisions.

{\it Acknowledgments.\/}
Fruitful discussions with A. Faessler and C. Fuchs are greatfully
acknowledged.
This work was supported in part by the Russian Federations
State Committee for Higher Education under grant E02-3.3-210,
Russian Fund of Basic Research (RFBR) under grant 03-02-16877, and 
Bundesministerium f\"ur Bildung und Forschung (BMBF) under contract
06T\"U986.

\newpage

\begin{figure}[t]
\centerline{
\includegraphics[width=190mm,height=190mm]{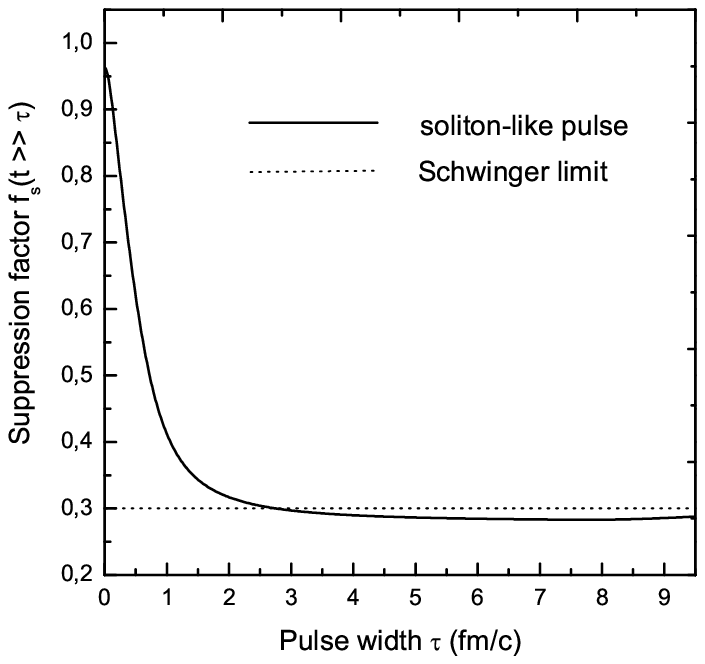}}
\caption{\lb{f1}
Pulse width dependence of the asymptotic value $f_s(t\gg\tau)$ of the
suppression factor $f_s$, the dotted line shows
the corresponding estimation from the Schwinger formula
\protect\re{eq3}.}
\end{figure}

\begin{figure}[t]
\centerline{
\includegraphics[width=190mm,height=190mm]{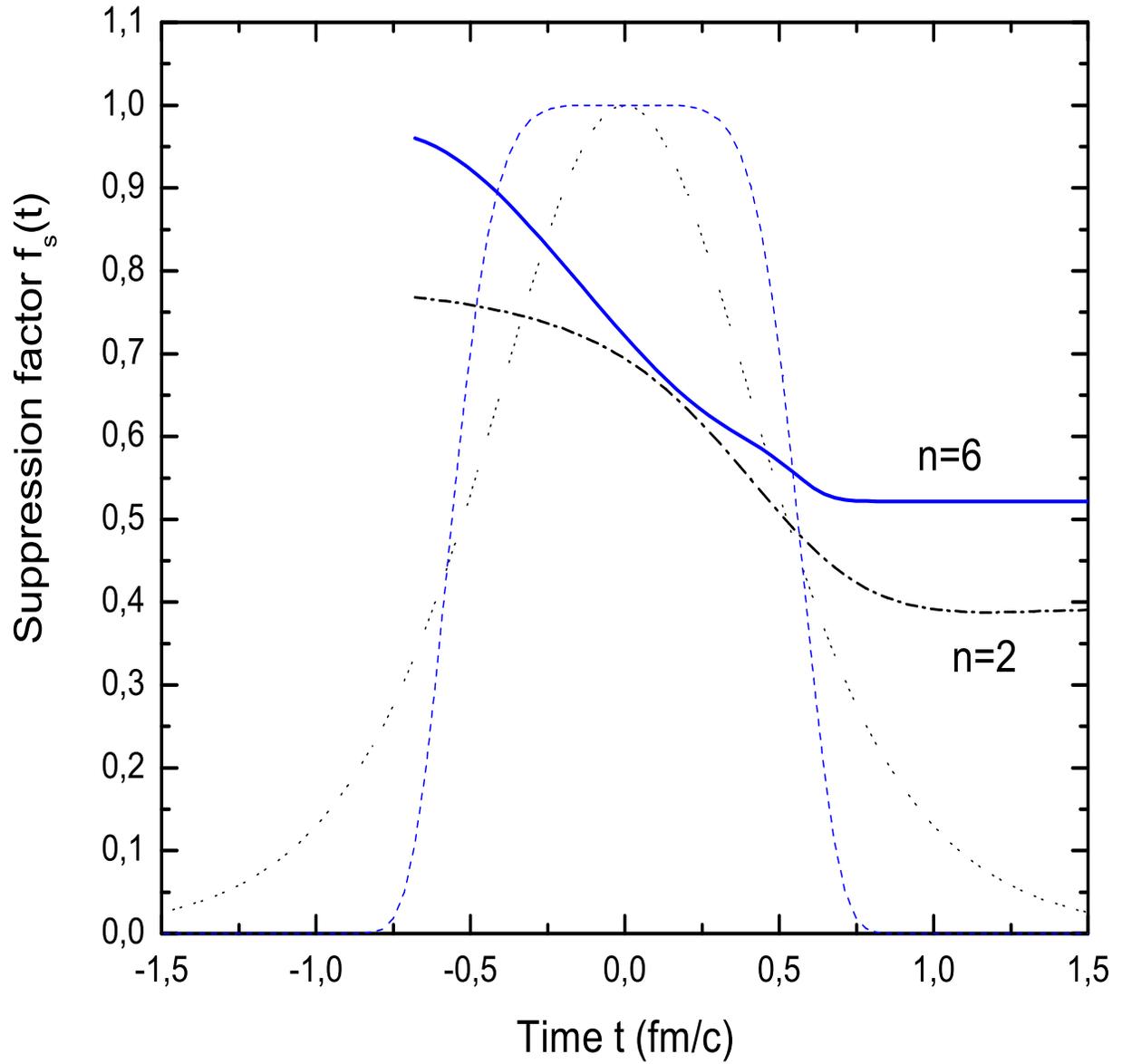}}
\caption{\lb{f2}
Time dependence of the suppression factor $f_s(t)$ for two types of field pulse,
given by Eq. \protect\re{gauss}, with the same width: the pulse with
$n=6$ (solid line) and the pulse with $n=2$ (dash-dotted line).
Dotted and dashed lines show the corresponding field profiles.}
\end{figure}

\begin{figure}[t]
\centerline{
\includegraphics[width=190mm,height=190mm]{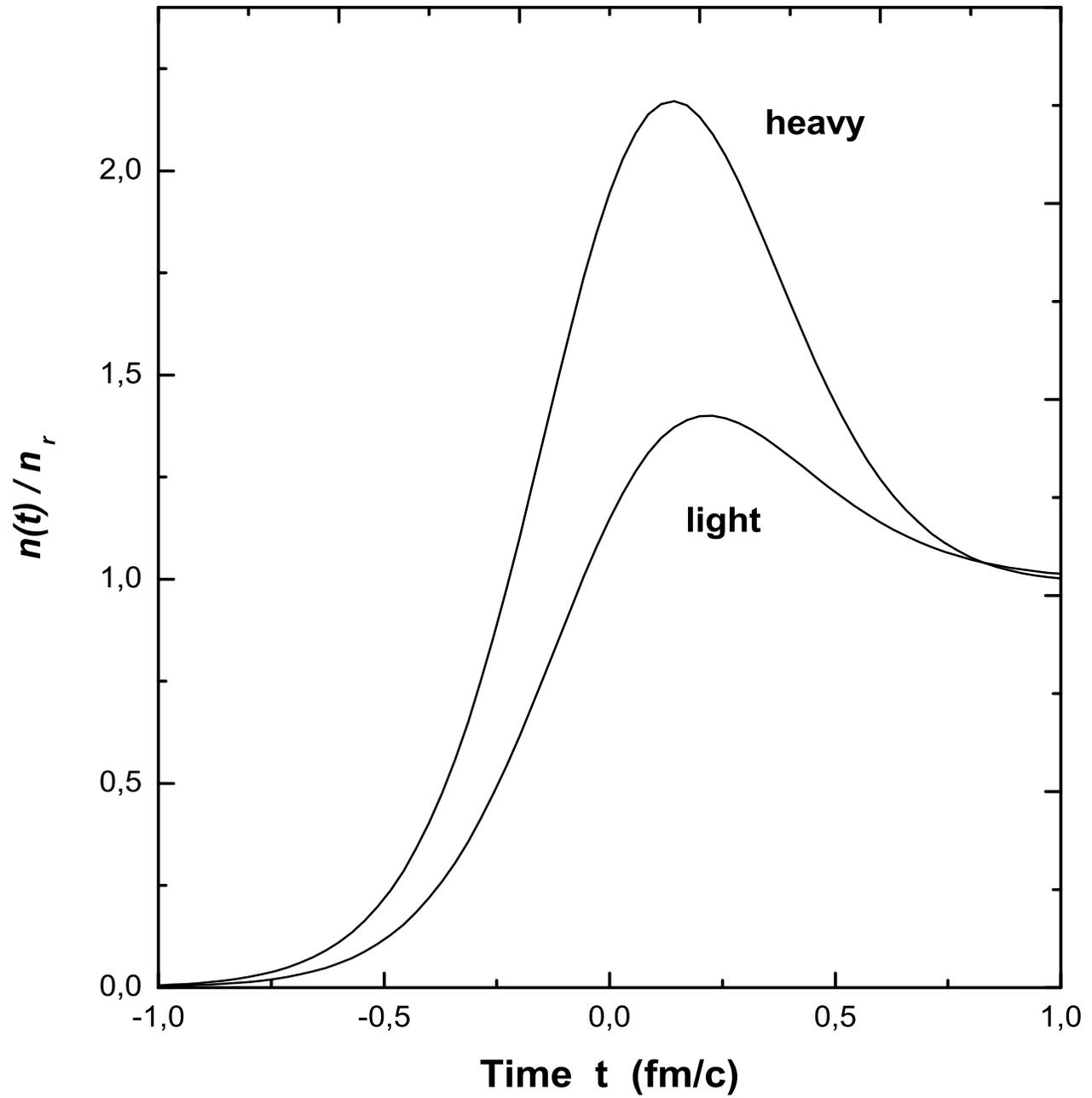}}
\caption{\lb{f3}
Time dependence of particle densities normalized on its final value
$n_r=n(t\gg\tau)$ for pulse \protect\re{eq27} with the
width $\tau_0 = 1$ fm/$c$.}
\end{figure}

\begin{figure}[t]
\centerline{
\includegraphics[width=190mm,height=190mm]{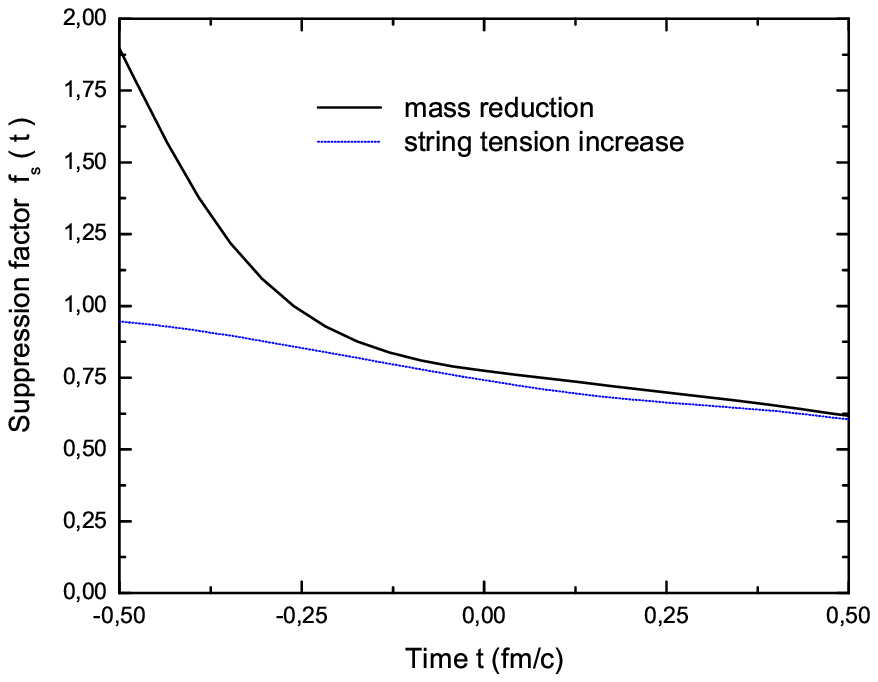}}
\caption{\lb{f4}
Comparison of two mechanisms of $f_s$ enhancement: the quark mass
reduction (solid line) and the string tension increase (dotted line).}
\end{figure}


\newpage

\begin{thebibliography}{99}

\bibitem{CNN79} A. Casher, H. Neuberger, and A. Nussinov,
Phys. Rev. D 20 (1979) 179; D 21 (1980) 1966.

\bibitem{lund} B. Andersson et al., Phys. Rep. 97 (1983) 31;
B.~Andersson, G.~Gustafson, and B.~Nielsson-Almqvist,
Nucl. Phys. B 281 (1987) 289.

\bibitem{Kaid99} A.B. Kaidalov,
Surveys in High Energy Phys. 13 (1999) 265;
A.B. Kaidalov and K.A. Ter-Martirosyan,
Phys. Lett. B 117 (1982) 247.

\bibitem{qgsm}
N.S.~Amelin and L.V.~Bravina, Sov. J. Nucl. Phys. 51 (1990) 133;
N.S.~Amelin, K.K.~Gudima, and V.D.~Toneev,
Sov. J. Nucl. Phys. 51 (1990) 1093.

\bibitem{venus} K. Werner, Phys. Rep. 232 (1993) 87.

\bibitem{urqmd} S.A. Bass et al., Prog. Part. Nucl. Phys. 41 (1998)
255; M. Bleicher et al., J. Phys. G 25 (1999) 1859.

\bibitem{schw} J. Schwinger, Phys. Rev. 82 (1951) 664.

\bibitem{low_str} F.~Antinori et al., WA97 Collaboration,
Nucl. Phys. A 661 (1999) 130c.

\bibitem{raf91} J.~Rafelski, Phys. Lett. B 62 (1991) 333.

\bibitem{bnk84} T.S. Biro, H.B. Nielsen, and J. Knoll,
Nucl. Phys. B 245 (1984) 449.

\bibitem{sor92} H. Sorge et al., Phys. Lett. B 289 (1992) 6.

\bibitem{am93} N.S. Amelin, M.A. Braun, and C. Pajares,
Phys. Lett. B 306 (1993) 312.

\bibitem{soff99} S. Soff et al., Phys. Lett. B 471 (1999) 89.

\bibitem{blei00} M. Bleicher et al., Phys. Rev. C 64 (2001) 011902.

\bibitem{WW88} R.-C. Wang and C.-Y. Wong, Phys. Rev. D 38 (1988) 348.

\bibitem{fbs}
D.V. Vinnik, A.V. Prozorkevich, S.A. Smolyansky et al.,
Eur. Phys. J. C 22 (2001) 341; Few-Body Syst. 32 (2002) 23.

\bibitem{grib} A.A. Grib, S.G. Mamaev, and V.M. Mostepanenko,
Vacuum Quantum Effects in Strong External Fields
(Friedmann Lab. Publ., St.-Petersburg, 1994).

\bibitem{gatoff} G. Gatoff, A.K. Kerman, and T. Matsui,
Phys. Rev. D 36 (1987) 114.

\bibitem{asakawa} M. Asakawa and T. Matsui,
Phys. Rev. D 43 (1991) 2871.

\bibitem{smol} S.A. Smolyansky et al., hep-ph/9712377;
S.M.~Schmidt et al., Int.~J.~Mod.~Phys. E 7 (1998) 709.

\bibitem{kluger98} Y. Kluger, E. Mottola, and J.M. Eisenberg,
Phys. Rev. D 58 (1998) 125015.

\bibitem{gan} A. Gangulu, P.K. Kaw, and J. Parikh,
Phys. Rev. D 48 (1993) R2983.

\bibitem{pop} S.V. Popov, JETP Lett. 74 (2001) 133.

\bibitem{rob} C.D. Roberts, S.M. Schmidt, and D.V. Vinnik,
Phys. Rev. Lett. 87 (2001) 193902; 89 (2002) 153901.

\bibitem{smol1} S.M. Schmidt et al., Phys. Rev. D 59 (1999) 094005;
J.C. Bloch et al., D 60 (1999) 1160011.

\bibitem{skok1} V.V. Skokov, S.A. Smolyansky, and V.D. Toneev,
hep-ph/0210099.

\bibitem{heinz} U. Heinz, AIP Conf. Proc. 602 (2001) 281.

\bibitem{oh} S. Ochs and U. Heinz, Ann. Phys. (NY) 266 (1998) 351.

\bibitem{qcd} A.V. Prozorkevich, S.A. Smolyansky, and S.V. Ilyin,
hep-ph/0301169.

\bibitem{mam} S.G. Mamaev and N.N. Trunov,
Sov. J. Nucl. Phys. 30 (1979) 677.

\bibitem{KES} Y.~Kluger, J.M.~Eisenberg, and B.~Svetitsky,
Int.~J.~Mod.~Phys. E 2 (1993) 333.

\bibitem{nar} N.B. Narozhnyi and A.I. Nikishov,
Sov. J. Nucl. Phys. 11 (1970) 596;
S.V. Popov, Sov. Phys. JETP 35 (1972) 659.

\bibitem{amel}
N.S. Amelin et al., Eur. Phys. J. C 22 (2001) 149.

\bibitem{cooper} F. Cooper, E. Mottola, and G.C. Nayak,
Phys. Lett. B 555 (2003) 181.

\bibitem{Bj83} J.D. Bjorken, Phys. Rev. D 27 (1983) 140.

\end{thebibliography}
\end{document}